# Identifying the Nano Interface Through Phase

Madhuri Mukhopadhyay[*]

*AI4ICPS, Indian Institute of Technology,*

*Kharagpur, West Bengal 721302, India*

*School of Mathematics and Natural Science, Chanakya University,*

*Bangalore, Karnataka 562129, India*

E-mail: mmukhopadhyay02@gmail.com, madhuri.m@chanakyauniversity.edu.in

**Abstract**

The quantum dots (QD) interface in solution can play significant roles in electron transfer dynamics for quantum dots-sensitized solar cells and different biological, environmental, and industrial systems. Here we predict an avenue to identify the contribution of quantum dots' interface-created static electric field, on the nonlinear optical response (NLO) due to four-wave mixing (FWM), especially for the nanoparticles where surface contribution is high. We implement a way to disentangle the FWM response in QDs, originating from the three incoming oscillating laser fields ($NLO_{oscillating}$) and a contribution ($NLO_{static}$) arising from the three oscillating laser fields and the static electric field caused by the interface. Advanced two-dimensional electronic spectroscopy (2DES) employs phase-resolved heterodyne techniques where FWM response is measured in a particular phase-matched direction and the response is distinctively phase sensitive. Theoretical analysis shows alteration in the interface can introduce phase variation in the $NLO_{static}$ signal resulting in distinct change in the 2D-spectra. Our studies establish a range of ionic strength ($10^{-6}M < x < 10^{-3}M$), which can be important to untwine, the usual NLO signal ($NLO_{oscillating}$) from the NLO ($NLO_{static}$) contributed by the interface of quantum dots. This analysis may open up the possibility to study the different kinds of dynamics occurring specifically in the interface and also will pave the path towards different ion interactions through phase change in 2D spectra and enormous scope will be employing deep learning-assisted phase recognition.

# Introduction

Quantum dots and nano-particles exhibit elevated surface-to-volume ratios, wherein the interfaces between these nano-particles and their surrounding environment exert crucial influences on their inherent physical and chemical characteristics.[1–4]

Especially the electron or proton transfer or the energy transfer dynamics at the interface can be influenced significantly depending on the nature of the interfaces. Understanding the nature and the role of the interfaces is fascinating to understand the overall ultra-fast dynamics of those systems.

Again, interfaces play a substantial role in diverse fields, such as drug interactions within biological membranes,[5,6] heterogeneous catalytic systems, paint industries, environmental and geoscientific studies,[7,8] interactions between water droplets and minerals, as well as photovoltaic device[9] applications. Non-linear spectroscopic techniques being non-invasive or non-destructive, provide a novel perspective to understand the properties of the hydrophilic and hydrophobic interface of nanoparticles, the molecular orientation and the distribution of molecules at particle surfaces in solution, interfacial structure of surfactants, the dynamics and transport properties.[10,11]

Over the past three decades, a wealth of literature has been dedicated to exploring the applications of non-linear spectroscopic studies in the realm of surfaces and interfaces. Commencing with conventional second harmonic and sum frequency generation techniques, the field has witnessed rapid advancements in novel methodologies for probing surface and interface properties.[12–18] These include cutting-edge techniques such as heterodyne detected two-dimensional vibrational sum frequency spectroscopy (VSFS), vibrational sum-frequency generation microscopy, as well as polarimetric angle-resolved second-harmonic scattering (AR-SHS), among others.[19–25] Shen et al.. demonstrated the significance and versatility of optical second harmonic generation and sum-frequency generation in surface and interface studies.

Furthermore, recent advancements highlight the utility of phase-sensitive sum-frequency vibrational spectroscopic techniques in obtaining both the real and imaginary components of the surface nonlinear spectral response. This approach enables the determination of water species orientations at the interface.[19,20,26–28] The flip-flopping behavior of interfacial water molecules, triggered by the opposite charge of surfactants, has been investigated by applying heterodyne-detected broadband vibrational sum frequency generation spectroscopy.[21] A recent in-depth study has shown how the flip of water molecules happens in the Stern layer with the change of pH compared to the diffuse layer.[29] This spectroscopic technique has also been utilized to capture the intricate spectra arising from air/water interfaces containing charged surfactants. The rearrangement of water molecules at the air/water interface, driven by the objective of minimizing surface free energy through the formation of hydrogen bonds with neighboring water molecules, has been extensively characterized through experimental and theoretical approaches.[30] Also, several SHG and SFH experiments were done to study solid liquid interfaces for different mineral oxides like silica, alumina, titania, etc.[31,32] The solid/liquid interfaces in colloidal quantum dots (QDs) play a pivotal role in diverse domains of natural sciences and industrial applications, encompassing fields such as photovoltaics[33,34] and medical applications.[35] In the realm of photovoltaics, a comprehensive understanding of efficient exciton transport, which directly impacts the overall efficiency of photovoltaic devices, heavily relies on the intricate nature of interactions and structures at the interfaces. Similarly, the establishment of a structure-function relationship of QDs at biological interfaces holds significant importance for the successful utilization of QDs in biomedical applications. Despite the availability of numerous techniques, ranging from surface second harmonic generation to heterodyne-detected phase-resolved two-dimensional vibrational sum frequency generation, for studying QDs with their high surface-to-volume ratio, these approaches primarily focus on the surface and interface regions.

However, they fail to provide comprehensive information when one seeks to investigate both the interface and the interface-to-bulk interactions within the same experimental setup. One example of a powerful technique is heterodyne-detected two-dimensional vibrational sum frequency spectroscopy (2D VSFS), which incorporates phase information to determine both the imaginary and real components of the second-order nonlinear susceptibility. This enables a more detailed analysis of molecular orientations based on the nature of vibrational modes. However, this technique lacks efficiency in establishing a direct connection between changes occurring at the interface and the bulk information.[23] In contrast, 2D electronic spectroscopy,[36–41] as a phase-resolved third-order nonlinear process, offers the advantage of a single experimental setup to efficiently provide comprehensive electronic information about the system, bridging the interface-to-bulk connection. With appropriate experimental design and meticulous data analysis, 2D electronic spectroscopy can complement VSFS by elucidating surface potentials, electronic environments, and charge separation processes across the entire system, extending from the interface to the bulk.

Standard techniques such as interface-sensitive SHG and VSFS have been employed to inves-tigate interface effects, capitalizing on the breaking of the center of symmetry at the interface. In contrast, the second-order non-linear technique, VSFS necessitates alterations in both dipole and polarizability, which predominantly occur at the interface. The key concept is that the contribution from the bulk solution is negligible, with the primary signal arising from the loss of inversion symmetry at the interface and the decaying electric field at the surface.[12–14,14–17,42]

Previously, it was believed that interface effects would be confined within a narrow region of 1-2 nm, in contrast to the non-linear coherence length, which could extend to tens of nanometers depending on the structural characteristics. However, if the electric field at the interface becomes sufficiently intense and comparable to the non-linear coherence length, interference effects may arise. In such cases, the non-linear optical signal at the interface is influenced not only by the surface potential but also by a screening factor, which encompasses the interface effects as a function of both the wave vector mismatch and the Debye length.[14,42,43]

Now, in the case of plasmonic quantum dots such as gold nanoparticles and graphene quantum dots, where the loosely bound surface electrons exhibit strong interactions with the electromagnetic radiation in NLO experiments, showcasing an overall excitonic behavior, we expect the primary contribution to NLO signals to arise from the surface, with a diffuse contribution extending from the surface to the bulk. Where SHG, SFG, VSFG, etc would limit our understanding to surface properties alone. However, by utilizing standard 2DES, we can overcome this limitation. 2DES enables the comprehensive investigation of both surface and bulk information in quantum dots, providing a holistic view of the entire system.

By monitoring the interface as a function of wave vector mismatch and the Debye length, 2DES allows us to capture the intricate dynamics occurring across the entire quantum dot system, from interface to bulk. Moreover, 2DES exhibits remarkable sensitivity to phase information, further enhancing its superiority in elucidating the interplay between surface and bulk phenomena. In the context of the interface, the presence of solvent molecules, which maintain a specific orientation depending on the surface potential, renders this region highly susceptible to polarization effects. The interface, in conjunction with the electrical double layer, gives rise to the establishment of a static electric field that creates an NLO response termed $NLO_{static}$. Our studies reflect, that beyond this interface-sensitive response, there exists an interface-insensitive response known as $NLO_{oscillating}$. The orientation of molecules at the interface fosters interference effects between NLO responses originating from different layers, resulting in a phase mismatch.

In our study, we aimed to investigate the interface effects as a function of both ionic strength and phase mismatch. It is important to note that the specific variations in interface effects may depend on the shape, size, and composition of quantum dots. Our findings reveal that within a certain range of ionic strength ($10^{-6}M < x < 10^{-3}M$), the $NLO_{static}$ signal exhibits a maximum, allowing it to be distinguished from the usual four-wave mixing signal, $NLO_{oscillating}$. 2DES provides control over phase information, can be employed to detect the $NLO_{static}$ signal. Depending on the amount of phase mismatch ($\Delta kz$) induced by the interface, the $NLO_{static}$ signal is expected to be spatially close to the $NLO_{oscillating}$ signal in the boxcar geometry for the rephasing pulse sequence.

Our study demonstrates distinct changes in phase and intensity in the $NLO_{static}$ 2DES spectra with variations in ionic strength. We also anticipate that for four-wave mixing (FWM), the main contributing signal is $NLO_{static}$, especially at low ionic strengths, as opposed to NLOoscillating. Therefore, our findings suggest that the search for the $NLO_{static}$ signal through 2DES experiments can provide explicit insights into the effects of the interface in this rapidly developing field. Thus combination of $NLO_{static}$ and $NLO_{oscillating}$ can give rise to from both interfaces to towards bulk information.

## Theoretical background

Now, the NLO interface field took shape by the fundamental work of Eisenthal and co-workers,[13,43,44] who directly linked the chemical environment of the interface as the static electric field at the interface and interpreted the second harmonic generation (SHG) signal from interfaces as consisting of the components χ(2) and the interfacial potential dependent third order component χ(3), as represented in equation 1). Which we can further interpret as equation 2) where $E_{oscillating}$ represents the $\chi(2)$ part resulting from the incoming oscillating laser fields and $E_{static}$ field resulting from the oscillating laser fields and the static electric field on the surface.

$$E_{sig} = \chi(2) + \chi(3)\Phi(0) \tag{1}$$

$$E_{sig} = E_{oscillating} + E_{static} \tag{2}$$

The static (DC) electric field $E_{DC}$, the reorientation, and the polarization of the solvent molecules contribute to the interface-dependent $\chi(3)$ term. Where the $\Phi_0$ is the surface potential. In those initial works $E_{DC}$ contribution was assumed to be due to the zero plane and the optical fields were considered to be independent of the depth of solvent from the surface.

Now, when a nanoparticle with the charged interface is immersed in an electrolyte solution, the ions in the solution interact and orient accordingly.[45-48] The static electric field due to charge interfaces is screened by the oriented counter ions in the solutions. Often these kinds of systems are modeled as electrical double layers (EDL).[46] Where the first layer of ions is adsorbed to the object and the next layer is formed by the Coulomb attractions of the counter ions which are loosely bound and diffusible. A typical Schematic of EDL is shown in Figure 1 a). Where the surface charge is often represented as the Surface potential.[49,50] Now the oriented solvent molecules near the interface are highly polarizable, and the effective potential decays depending on the length scale of the orientation of those molecules. The interfacial double layer has been described using different models such as the Helmholtz model, Gouy Chapman model, Stern model, etc.[51,52] The electrostatic effect is often considered to decrease exponentially. The charge distributions are generally described by the Poisson-Boltzmann equation. In the Gouy Chapman model, the Poisson Boltzmann equation is solved for the planar surface and the surface potential varies with the ionic strength of the electrolyte in the solution, as given by (equation 3). Gouy Chapman model can be used for bigger sizes of quantum dots as it will be valid to charged curve surface when $\kappa r > 15$, where r is the radius of the particle or quantum dot.[42]

$$\Phi(0) = \frac{2k_BT}{Z_v e}\sinh^{-1}\sqrt{\frac{\sigma}{\frac{2}{\pi}\varepsilon k_B T C}} \qquad (3)$$

Where, $Z_v e$ is the valence of the ions, C electrolyte concentration, $\sigma$ surface charge density, $\varepsilon$ is the dielectric constant of the bulk solvent.

Hence the effective static electric field is given by $E_{DC}(z) = -\frac{d}{dz}\Phi(z)$ where $\Phi(z) = \Phi_0 e^{-\kappa z}$. Where $\kappa^{-1} = \lambda =$ Debye length.

Again the oriented solvent molecules near the surface having particular alignment and the decaying electrostatic potential give rise to interference effects on the light interactions from the different layers of interface, modulating the overall coherence length of the processes (Fig. 1 b). So the optical field is dependent on the depth (z) from the surface and the phase mismatch for the nlo processes due to the optical dispersion and is given by $\Delta k_z = |k_{1z} + k_{2z} - k_{sig,z}|$. Where $k = \frac{\omega_i}{c}\sqrt{n(\omega)^2 - \sin(\theta)^2}$ [42,53]

Recently people have tried to untwine the dynamic and the static part with the variation of both $\kappa$ and the phase of mismatch $\Delta k_z$. Ohono *etal* have employed the Heterodyne detected SHG having control over the phase mismatch and applied different ionic strength to extract the static contribution as a function of the phase angle $\phi$, as given by equation 6).[54] The static contribution to the total nonlinear polarization SHG/SFG signal is controlled by both the electrical developments and phase-matching consequences, which is given below by Ohono *etal*.[18,54]

$$SFG_{static} = \varepsilon_0 \int_0^\infty \chi^{(3)} E_1(\omega_1, k_1) E_2(\omega_2, k_2) E_{DC}(z) e^{i\Delta k_z z} dz \qquad (4)$$

The integration by parts gives to

$$SFG_{static} = \varepsilon_0 \chi^{(3)} E_1(\omega_1, k_1) E_2(\omega_2, k_2) \Phi(0) \frac{\kappa}{\kappa - i\Delta k_z} \qquad (5)$$

Separating the real and the imaginary part signifies that the static field contributes to both the absorptive and the dispersive part of the signal.

$$\frac{\kappa}{\kappa - i\Delta k_z}\chi^{(3)} = \frac{\kappa^2}{\kappa^2 + \Delta k_z^2}\chi^{(3)} + i\frac{\kappa\Delta k_z}{\kappa^2 + \Delta k_z^2}\chi^{(3)} \quad (6)$$

Hence one would expect the change in the spectral line shape controlled by the Debye length as well as the phase mismatch. The phase angle of this process is given by $\phi = arctan(\frac{\Delta k_z}{\kappa})$

Thus previous studies[18,52,54] already show total second harmonic signal is going to be the contribution from the $\chi^{(2)}$ as well as $\chi^{(3)}$ part due to the contribution of ionic potentials as well the phase contribution as given by equation 7).

$$E_{SHG} \propto \chi^{(2)} + \chi^{(3)}\Phi(0)cos(\phi)e^{i\phi} \quad (7)$$

## Results and Discussions

Fig 1 c) presents the quantitative variation of the Surface potential and the inverse Debye length with the change of ionic strength of the solution.

Fig 1. d) shows the variation of phase $\phi$ as a function of both ionic strength and the phase mismatch. This reflects that phase change can be employed as a significant parameter to understand the interface.

As the advanced two-dimensional electronic spectroscopy (2DES)[36-40] employs phase resolved heterodyne techniques having control over the phase information we consider this procedure to understand the interface of nano/microparticles. Fig 2 a) shows the schematic of the pulse sequences for heterodyne detected four-wave mixing, where the signal is collected in rephasing phase-matched direction interfering with a local oscillator(LO).[55]

The 3rd-order nonlinear polarization of an isotropic system is given as the time evolution of the density matrix by the third-order perturbation of three laser fields resulting in the four-wave mixing signal as given by equation 8 and shown schematically in Fig 2)b) through density matrix descriptions.[56,57] Fig 2)c) Shows the typical 2DES spectra for coupled oscillators, where on the left side there are corresponding linear spectra. Where practically $\omega_1$ is the pump frequency and $\omega_3$ is the probe frequency and are obtained by effectively Fourier transform of t1 and t3 time delays. Fig. 2)d) is the wave vector for rephasing the pulse sequence. This clearly reflects that we can expect spatial shift for ($nlo_{oscillating}$) and ($nlo_{static}$) depending on the extent of phase mismatch ($\Delta k_z$).

$$P^{(3)}_{FWM} = NLO_{oscillating} = -\left(\frac{i}{\hbar}\right)^n \int_0^\infty dt_3 \int_0^\infty dt_2 \int_0^\infty dt_1 \, E(t-t_3) E(t-t_3-t_2) E(t-t_3-t_2-t_1) \quad (8)$$

$$\langle \mu(t_3+t_2+t_1)[\mu(t_2+t_1),[\mu(t_1),[\mu(0),\rho(-\infty)]]\rangle$$

Here pertinent to mention that there are some standard assumptions and experimental tricks to handle the equations reasonably like the dipole approximations are valid, there is an ordering of the pulse sequence, the rotating wave approximations are taken into account and the pulses are considered to be ultra-fast so that can be assumed to be delta functions.[56,57] Hence for different phase matching conditions, different response functions (Equation 9) can be obtained which gives the spectral details of the FWM signal.

$$S_3 = -\left(\frac{i}{\hbar}\right)^n \langle \mu(t_3+t_2+t_1)[\mu(t_2+t_1),[\mu(t_1),[\mu(0),\rho(-\infty)]]\rangle \quad (9)$$

Now the colloidal nano/micro particle systems can have surface-enhanced properties. In that case, the interfacial potential will contribute to the total nonlinear polarization. So we can consider the total nonlinear polarization as a sum of both from oscillating contribution and the static electric field contribution from the surface, similar to equation 2.

$$P^{(3)}_{total} = P^{(3)}_{oscillating} + P^{(3)}_{static} = NLO_{oscillating} + NLO_{static} = NLO_{total} \qquad (10)$$

Where the *NLO$_{oscillating}$* part will be the same as equation 8 and *NLO$_{static}$*, the static part will be contributed by the interfacial potential and the phase mismatch due to the dispersion caused by the interfacial distribution of the solvents.

Hence the *NLO$_{static}$* can be defined as the signal due to the three incoming laser fields and the static electric field at the interface and also the contributions from the phase mismatching.

$$NLO_{static} = -\left(\frac{i}{\hbar}\right)^n \int_0^\infty dt_3 \int_0^\infty dt_2 \int_0^\infty dt_1\, E(t-t_3)E(t-t_3-t_2)E(t-t_3-t_2-t_1) \int_0^\infty E_{DC}(z)e^{i\Delta k_z z}dz$$
$$\langle \mu(t_3+t_2+t_1)[\mu(t_2+t_1),[\mu(t_1),[\mu(0),\rho(-\infty)]]]\rangle \qquad (11)$$

If we write the interfacial contribution in terms of response function it will be given by equations 11)

$$NLO_{static} = 3S \int_0^\infty E_{DC}(z) e^{i\Delta k_z z} dz \qquad (12)$$

$$NLO_{static} = S_3 \int_0^\infty -\frac{d}{dz}\Phi_0 e^{-\kappa z} e^{i\Delta k_z z} dz \qquad (13)$$

Integrating by parts, (see SI for the details)

$$NLO_{static} = S_3 \left[ \Phi_0 + i\Delta k_z \int_0^\infty \Phi(z) e^{i\Delta k_z z} dz \right] \qquad (14)$$

$$NLO_{static} = S_3\ \Phi_0 \frac{\kappa}{\kappa - i\Delta k_z} \tag{15}$$

The total contribution of the 3rd-order nonlinear signal is given by

$$NLO_{Total} = NLO_{oscillating} + S_3\Phi(0)cos(\phi)e^{i\phi} \tag{16}$$

Where $\phi = arctan(\Delta k_z\lambda)$ and $\Phi_0$ is given by equation 3. See SI for the details.

Fig 3 a) and b) Shows the change of absolute and imaginary parts of the $nlo_{static}$ response as a function surface potential only. We observe a steady increase of the $nlo_{static}$ response with the increase of the surface potential, without any peak shift, whereas in Fig 3 c) and d) we notice both peak and intensity change with the change of phase angle $\phi$ which is a function of both Debye length and phase mismatch.

Fig 4 a) to e) systematically differentiate the effects of different parameters on $nlo_{static}$ response. With a variation of surface potential ($\Phi_0$) the $nlo_{static}$ increase linearly a), whereas the $nlo_{static}$ response decreases exponentially with the increase of the distance (Z) from the surface charge b), $nlo_{static}$ initially increases then decreases following an arctan curve for the change of phase mismatch c). With the change of Debye length Fig 4 d) the $nlo_{static}$ decreases exponentially initially, then varies linearly showing little variation with the change of Debye length. Fig 4 e) reflects the change in $nlo_{static}$ with the variation of phase angle. Which clearly shows a periodic alternation of $nlo_{static}$ with the phase angle $\phi$, for different values of surface potential. It also reflects that for the system where the surface potential is negligible even the change of phase mismatch can not significantly give a high $nlo_{static}$ response. Thus if phase mismatch is zero but the surface potential is not still we will get a strong $nlo_{static}$ signal but the reverse is not true.

In Fig. 4 f) the left side shows the variation of $nlo_{static}$ with the alternation of ionic strength. The yellow curve, where it is a function of both the phase factor $\frac{\kappa}{\kappa - i\Delta k_z}$ and the surface potential($\Phi_0$) increases and then decreases again. Where the red curve shows the change of $nlo_{static}$ as a function of ionic strength when the surface potential contributes only and it shows a sharp decrease initially and then a slow decrease to a constant for high ionic strength. Interestingly the red and the yellow curves merge at high ionic strength. The blue curve shows the variation of the phase factor with the ionic strength, which increases rapidly in the low ionic strength and then remains constant with the increase of the ionic strength, as shown in the right axis. So Fig. 4 f) significantly shows how the overall $nlo_{static}$ is controlled by both the phase factor and the surface potential part.

In Fig. 4 g) we plot $nlo_{static}$ as a function of the log of ionic strength for different phase mismatch $\Delta k_z$, which shows for the range of ionic strength ($10^{-6}M < x < 10^{-3}M$) the $nlo_{static}$ increases and then again decreases with the increase of the ionic strength.

Fig. 5 is the 2DES spectra for different ionic strengths, 0.000001 M, 0.00005 M, and 0.005M. There is distinctly phase change and also a change in the intensity with the variation of the ionic strength. As at low ionic strength phase angle approaches $\pi/2$ hence as a whole signal decreases. At intermediate ionic strength, intensity is highest with distinct phases compared to higher concentrations.Hence it clearly reflects that at intermediate ionic strength (($10^{-6}M < x < 10^{-3}M$)), $NLO_{static}$ is distinctly different from $NLO_{oscillating}$. Hence we propose 2DES experiment can disentangle $NLO_{static}$ from $NLO_{oscillating}$ in this ionic strength range.

SI shows the 2DES spectra for $NLO_{oscillating}$ which is independent of ionic strength and is the same phase as $NLO_{static}$ of high ionic strength but with much less intensity. We also anticipate $NLO_{static}$ is the main contributing part of $NLO_{Total}$ signal. At high ionic strength, it acts like $NLO_{oscillating}$, and we may not distinguish them separately.

Also, we suggest the likeliness of spectral broadening in the $NLO_{static}$ at low ionic strength. As shown in S.I. We also bring up the plausibility of other kinds of potentials and the necessity of profound research that could be interesting for H-bonded solvent molecules at the interface or because of surface passivating ligands, etc.

# Conclusion

For systems like quantum dots where the surface-to-volume ratio is high, the effects of interface could be imperative. Now, the solvent molecules remain oriented in the interface depending on the surface potential, making the region sensitive towards polarization. The surface potential along with the electrical double layer at the interface develops a static electric field which produces the *NLO$_{static}$* response. Hence we may expect two types of responses, *NLO$_{oscillating}$*, which is insensitive to the interface, and *NLO$_{static}$* sensitive to the interface. The oriented molecules in the interface induce interference between the generated NLO response from different layers effectively causing a phase mismatch also. In our studies, we tried to find out the effect of the interface as a function of both ionic strength and phase mismatch in general. However, there could be particular variations in the interface effects with the change of shape, size, and composition of quantum dots. Our study reveals that for a range of ionic strength ($10^{-6}M < x < 10^{-3}M$), *NLO$_{static}$* takes maxima when it can be dis-entangle from the usual four-wave mixing signal *NLO$_{oscillating}$*. The advanced 2DES technique having control over the phase information can be used to detect the *NLO$_{static}$* signal as well *NLO$_{oscillating}$*, giving both interface and the bulk information for the system. Depending on the amount of phase mismatch $\Delta k_z$ produced by the interface, one can expect the *NLO$_{static}$* spatially close to the *NLO$_{oscillating}$* signal in the box car geometry for the rephasing pulse sequence. The studies reflect distinct phase and intensity changes in the *NLO$_{static}$* 2DES spectra when there is a variation in ionic strength. We also anticipate for FWM the main contributing signal is *NLO$_{static}$* especially in the low ionic strength, rather than *NLO$_{oscillating}$*. The phase change originated due to *NLO$_{static}$* contribution will provide the opportunity to identify different kinds of interface and the nature of the ions in the interface. We also foresee the use of deep learning in phase recognition will further enhance the field.

# Acknowledgement

MM acknowledges Indian Institute of Science for research facilities and funding. MM sincerely acknowledges Prof. P.K.Das, and his lab members of IPC, IISc for useful discussions and also thanks AI4ICPS, IIT Kharagpur for funding.

# Supporting Information Available

Detailed derivation of equations 11-17, discussion of electrical double layer and Debye length, Possibility of different kinds of interface potential, effects of the interface on the line shape function, relaxation, and the coupled phonon modes to the oscillators. Details of simulation methods and codes are also given.

The following files are available free of charge.

- 1. Detail derivation of equations 11-17

- 2. Discussion of electrical double layer and Debye length and supplementary figures

- 3. Effects of dephasing

- 4. Possibilities of different potentials

# TOC Graphic

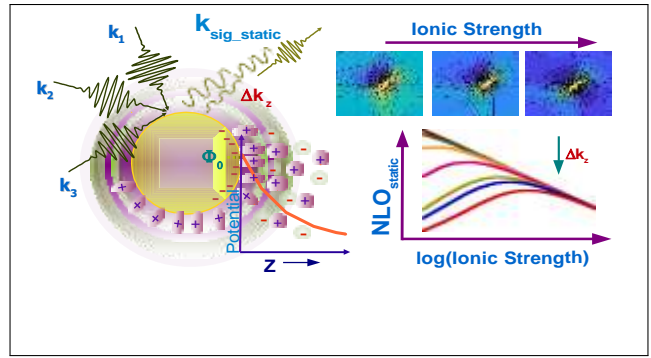

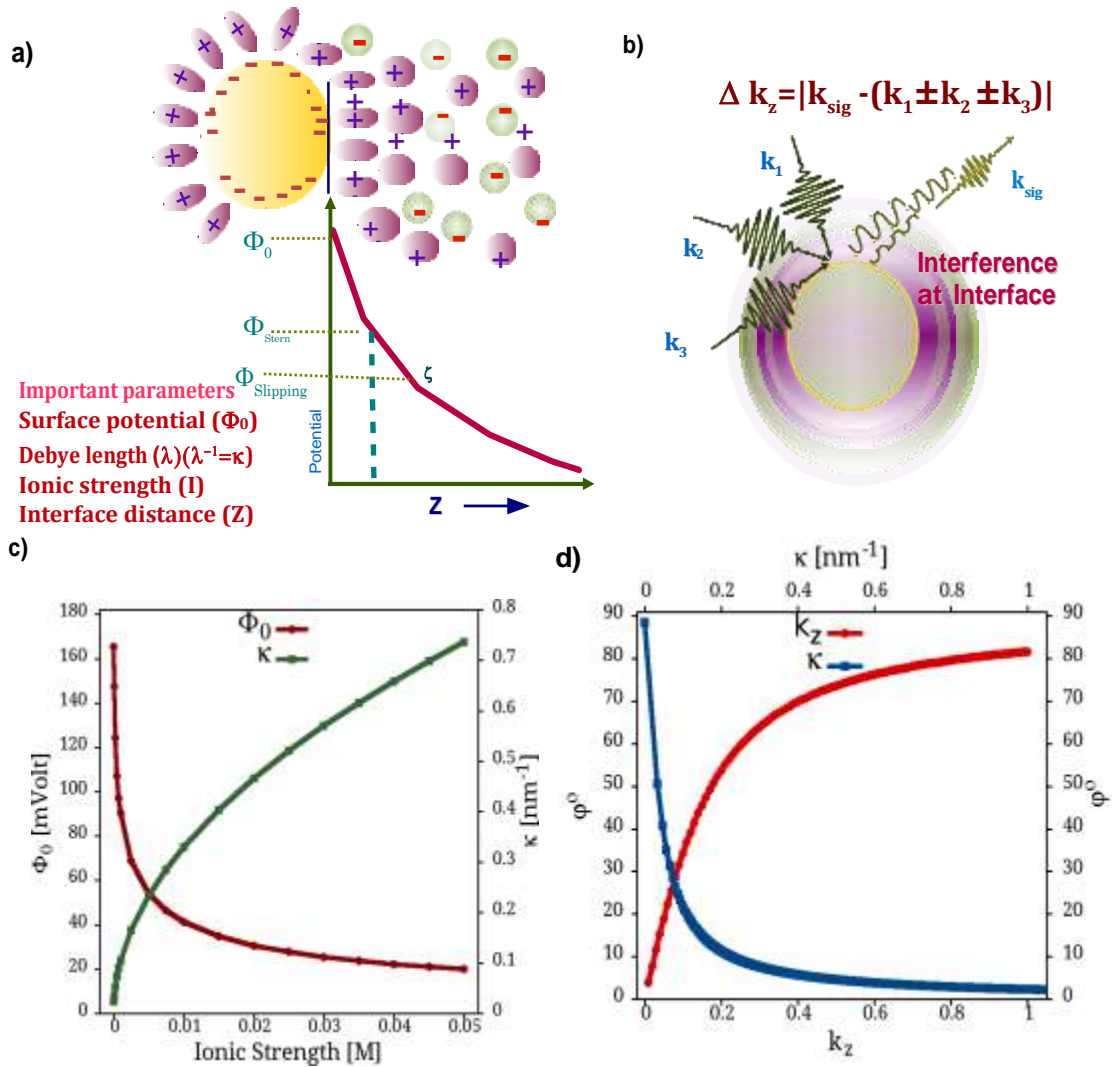

Figure 1: a) A simple schematic of the quantum dot interface in solution, the electric double layer, and the associate potentials as a function of Z, the distance from the surface. $\Phi_0$ is the surface potential b)Interface not being isotropic can cause interference in the signal generated from different layers resulting in phase mismatch ($\Delta k_z$). c)Variation of Surface potential, and inverse Debye length ($\kappa = \lambda^{-1}$) with respect to ionic strength d)Variation of Phase angle ($\phi$) with respect to phase mismatch and inverse Debye length.

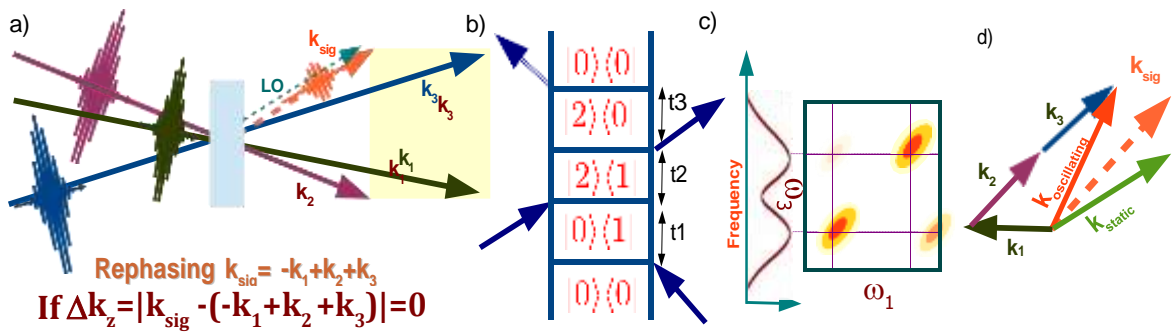

Figure 2: a) Schematic for the four waves mixing in rephasing geometry, for 2-dimensional electronic spectroscopy technique, where the signal is heterodyne detected using a local oscillator (LO). b) The double-sided Feynman diagram for rephasing pulse sequence. c) Cartoon of 2-dimensional electronic spectra, in comparison with its linear spectra for a coupled oscillator. d) Wave vector for rephasing geometry. In isotropic media $k_{Static}$ will vanish and $k_{dynamic}=k_{signal}$ but at interface $k_{static}$ may not be zero and can contribute to the resultant signal or it can create $P_{static}$ signal in the direction depending on ($\Delta k_z$)

.

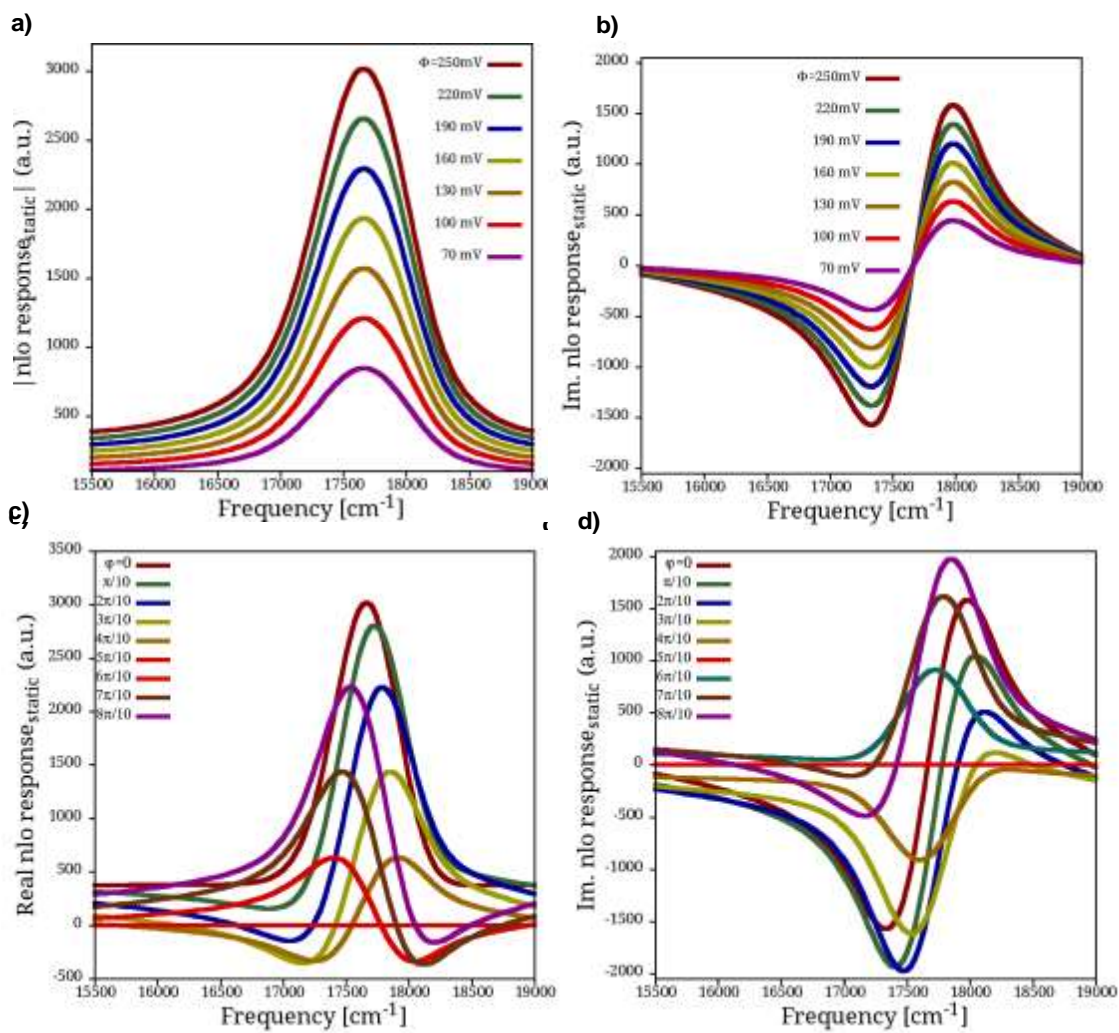

Figure 3: a), b) Absolute and Imaginary nlo response due to static component with a variation the surface potential only, without any phase effects. c), d)Real and Imaginary nlo response due to static component with variation the phase factor

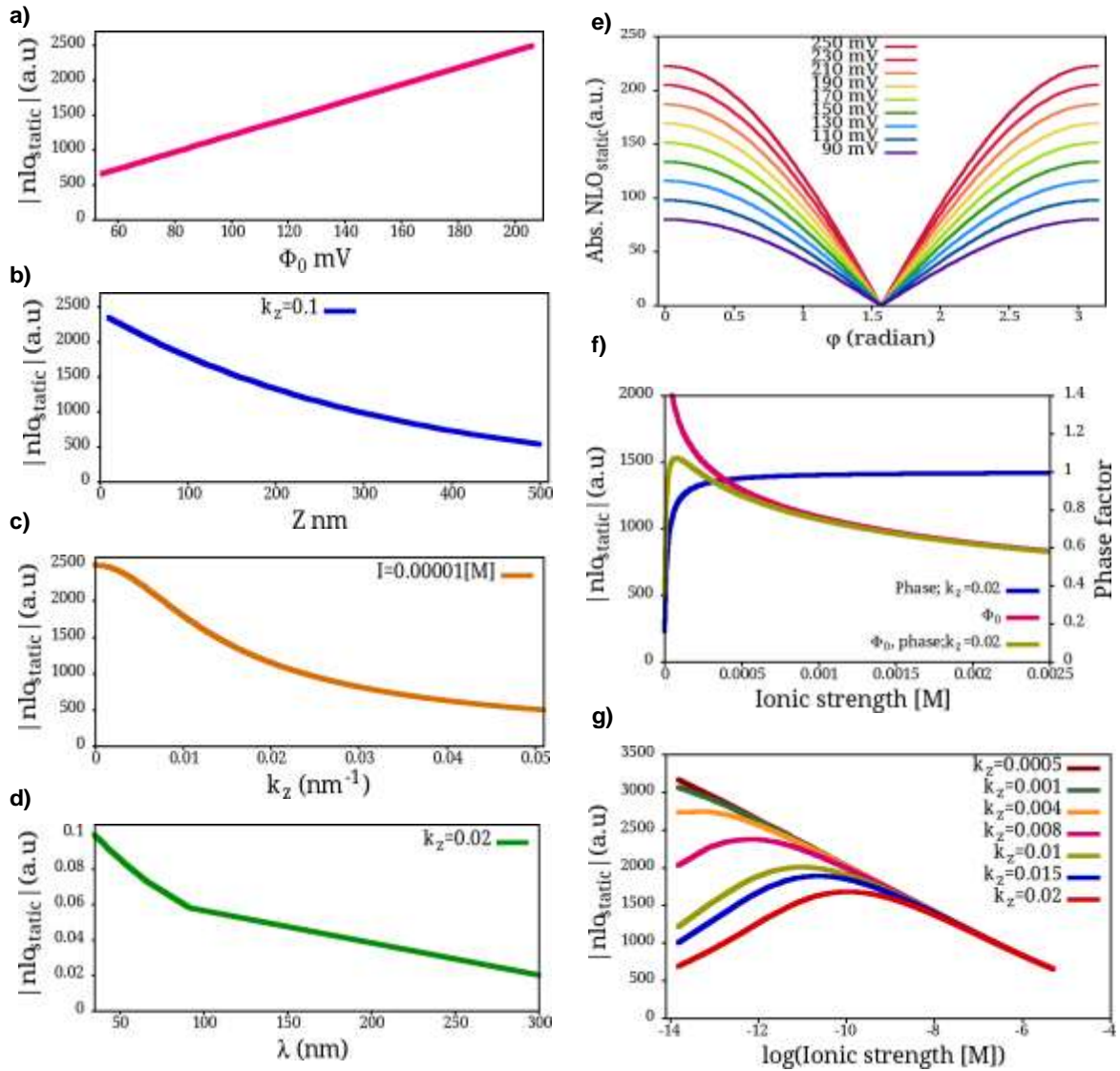

Figure 4: a),b),c),d) Absolute of nlo static part as a function of surface potential, distance from the surface, phase mismatch, and Debye length respectively. e)Variation of the absolute four-wave mixing signal due to the static part with the change of Ionic strength in the yellow-green curve. The Red curve shows the surface potential contribution and the blue part due to the phase factor. f) Absolute $Nlo_{Static}$ as a function of phase $\phi$ and Surface potential. g) Change of Absolute $Nlo_{Static}$ due to

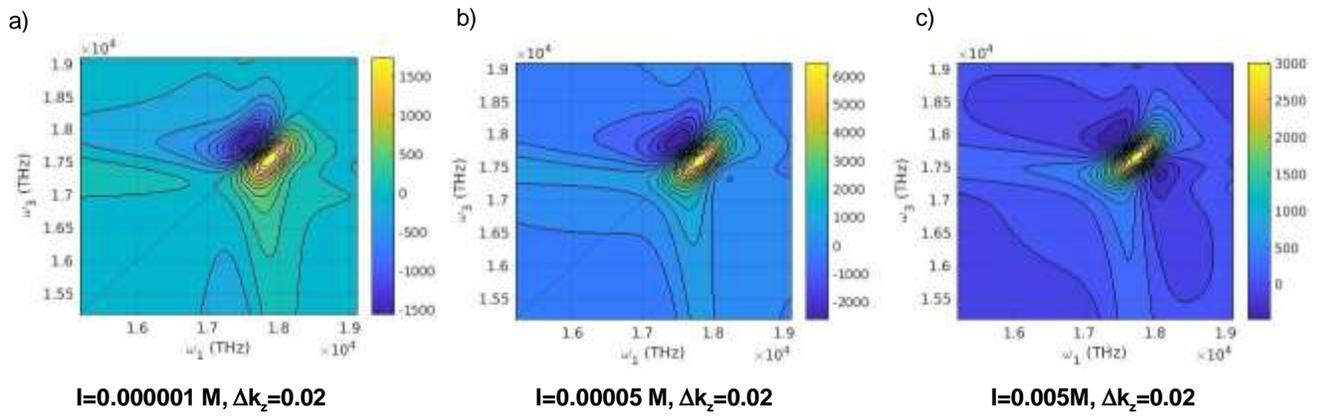

Figure 5: Two-dimensional four-wave mixing signal for the static part with the variation of ionic strength a)0.000001 M, b)0.00005 M, and c) 0.005M. Which clearly shows both phase and intensity change. As at low ionic strength, phase $\phi$ goes to $\pi/2$ hence as a whole signal decreases. At intermediate ionic strength intensity is high with distinct phase compared to higher concentration